\documentclass[onecolumn,showpacs,preprintnumbers,amsmath,amssymb]{revtex4}

\usepackage{graphicx}
\usepackage{dcolumn}
\usepackage{bm}

\begin{document}

\preprint{APS/123-QED}

\title{A Lindbladian From Feynman-Vernon}

\author{Jose A. Magpantay}

\affiliation{Quezon City, Philippines}
\email{jose.magpantay11@gmail.com}
\date{\today}

\begin{abstract}
The effective dynamics of a system interacting with a bath or environment is presented in two ways, (1) the (LGKS) replacement of the von Neuman equation for the density matrix and (2) the Feynman-Vernon path integral derivation, by integrating out the bath degrees of freedom, to arrive at the propagator of the system's density matrix. In this paper I connect the two methods by deriving a Lindbladian in a mechanical example, a point particle interacting with a bath of harmonic oscillators, previously considered by Feynman and Vernon (FV) and expounded on later by Caldeira and Leggett (CL). But the (FV)/(CL) results only in non-Markov effect, memory terms from the bath interaction. To derive a Lindbladian, I changed the interaction term they considered to take into account the point particle interacting with the bath harmonic oscillators to something more realistic. From the resulting path-integral expression of the system's propagator for the density matrix, the Lindbladian and the non-Markov term are read for this simple problem. I also point out the causes of these terms, the Markov Lindbladian from the very local interaction of the point particle with the classical solutions of the harmonic oscillator and the non-Markov term from the global interaction of the point particle with the fluctuations of the classical solutions. 
\end{abstract}

\pacs{Valid PACS appear here}
\maketitle

1. There are two ways to deal with the quantum statistics of open systems, the first and derived earlier using path-integral techniques is the Feynman-Vernon (FV) theory \cite{Feynman} and the second derived more than a decade later is what became known as the Lindblad equation \cite{Lindblad}, which was also independently derived at about the same time by Gorini, Kossakowski and Sudarshan (GKS) \cite{Gorini}, from hereon will be referred to as the (LGKS) approach. The (LGKS) approach derives the replacement of the von Neuman equation for the density operator/matrix (see any statistical mechanics book, for example  \cite{Reichl}), under the general conditions of Hermiticity, Markov process and complete positivity. The Feynman-Vernon starts from the system-bath interaction and derive, using the path-integral method, the system's density matrix at any time t by integrating out the environment/bath degrees of freedom. However, still at this time there has been no paper that relates (LGKS) from (FV). The link is crucial for it will provide how the terms that appear in the (LGKS), the Hamiltonian and the Lindblad operator, are given in terms of the system-bath dynamics. But in a certain sense, the link does not seem to be warranted for the simple reason that (FV), being a sum over histories approach would naturally be dominated by memory effects, thus non-Markov, while the (LGKS) is limited by the Markov process condition, along with two other conditions. 

Regardless of the above ideas, it is interesting to find out if the Lindblad equation can be derived from the Feynman-Vernon path-integral approach to quantum statistical mechanics. Maybe even in doing the sum over histories of the bath, there may be non memory effects that can be isolated. But most likely it depends on the interaction between the bath and the system. The (FV)/(CL) point particle interacting with a bath of harmonic oscillators clearly show the dominance of memory effects, the only path-integral result for this problem is what I called the Lindblad Plus  \cite{Magpantay}, the Plus represents the memory effect terms. 

2. I begin by summarizing known ideas about von Neuman type equations in statistical mechanics. First, for a closed system, described by a Hamiltonian $ H(x,p) $, the von Neuman equation for the density operator  $ \rho $ is given by, see for example \cite{Reichl} 
\begin{equation}\label{1}
i\hbar \dfrac{\partial \rho}{\partial t} = \left[ H, \rho \right] .
\end{equation}
The solution to equation (1) 
\begin{equation}\label{2}
\rho (t) = \exp {(-\frac{i}{\hbar} H t) } \rho (0) \exp {(+\frac{i}{\hbar} H t) },
\end{equation}
from which we get the time evolution of the density matrix $ \rho (x,y; t) $ as
\begin{equation}\label{3}
\begin{split}
\rho (x,y; t)& = \int dx'dy' K(x,x'; t) \rho (x',y'; 0) K^{*}(y',y; t)\\
                     & = \int dx'dy'J(x,y;x',y'; t) \rho(x',y'; 0).
\end{split}                      
\end{equation}
The $ K(x,x'; t) $ is the quantum propagator given by the path-integrals
\begin{equation}\label{4}
\begin{split}
K(x,x'; t)& = \int_{end pts} (d\tilde{x}) \exp {\frac{i}{\hbar} \int_{0}^{t} dt' L(\tilde{x},\dot{\tilde{x}})}\\
               & = \int_{end pts} (d\tilde{x})(dp_{x}) \exp {\frac{i}{\hbar} \int_{0}^{t} dt' \left[ p_{x}\dot{\tilde{x}} - H(\tilde{x},p_{x}) \right]},
\end{split}
\end{equation}
and $ J(x,y;x',y'; t) $ is the density matrix propagator and is given by
\begin{equation}\label{5}
\begin{split}
J(x,y;x',y'; t)& = K(x,x'; t) K^{*}(y',y; t)\\
                        & = \int_{end points} (d\tilde{x}) (d\tilde{y}) \exp {\frac{i}{\hbar} \int_{0}^{t} dt' \left[ L(\tilde{x},\dot{\tilde{x}}) - L(\tilde{y},\dot{\tilde{y}}) \right]} \\
                        & = \int_{end points} (d\tilde{x}) (d\tilde{y}) (dp_{x}) (dp_{y}) \exp {\frac{i}{\hbar} \int_{0}^{t} dt' \left( \left[ p_{x}\dot{\tilde{x}} - H(\tilde{x},p_{x}) \right] - \left[ p_{y}\dot{\tilde{y}} - H(\tilde{y},p_{y}) \right] \right)}.
\end{split}
\end{equation}
   
For an open system, i.e., a system interacting with a bath/environment, there are two major approaches - the Feynman-Vernon theory and the Lindblad and Gorini - Kossakowski - Sudarshan theory (LGKS). The LGKS provides the analogous von Neumann type of equation for the system under conditions of (1) Markov process, (2) Hermiticity and (3) complete positivity and it is given by
\begin{equation}\label{6}
\begin{split}
i\hbar \dfrac{\partial \rho}{\partial t}& = [H,\rho] + i\textbf{L} \rho \textbf{L}^{\dagger} - \frac{i}{2}\left( \textbf{L}^{\dagger}\textbf{L}\rho + \rho \textbf{L}^{\dagger}\textbf{L} \right) \\
                                                                     & = H_{eff} \rho - \rho H^{\dagger}_{eff} + i \textit{L} \rho \textit{L}^{\dagger},
\end{split}
\end{equation}
where $ \textbf{L} $ is the LIndbladian and $ H_{eff} $ is given by
\begin{equation}\label{7}
H_{eff} = H - \frac{i}{2} \textbf{L}^{\dagger}\textbf{L}
\end{equation}
This form of the LGKS equation is due to Struntz \cite{Struntz}, which provided the path integral for the time differential evolution operator of the system's density operator, the analogue of equations (3) and (4) for a closed system.  This form is significant for it will guide the inclusion of memory effects, the non-Markovian term, which must be present in general for a system interacting with a bath/environment for an extended period of time.  The path-integral version of equation (6) is given in \cite{Struntz} as
\begin{equation}\label{8}
\rho(x,y; t) = \int dx'dy' J_{L}(x,y; x',y'; t) \rho(x',y'; 0), 
\end{equation}
where the four-point plus time function $ J_{L} $ is the propagator for the density matrix of a Lindblad system and it is given by the system's coordinates and momenta path-integrals
\begin{equation}\label{9}
\begin{split}
J_{L}(x,y; x',y'; t)&  = \int_{end points} (d\tilde{x}) (d\tilde{y}) (dp_{x}) (dp_{y})  \exp \frac{i}{\hbar} \int_{0}^{t} dt' \big\lbrace \left[ p_{x}\cdot\dot{\tilde{x}} - H(\tilde{x},p_{x}) \right] \\
 &\quad - \left[ p_{y}\cdot \dot{\tilde{y}} - H(\tilde{y},p_{y}) \right] -i \left[ (\textbf{L}^{\dagger}\textbf{L})(\tilde{x},p_{x}) + (\textbf{L}^{\dagger}\textbf{L})(\tilde{y},p_{y}) + \textbf{L}(\tilde{x},p_{x})\textbf{L}^{\dagger}(\tilde{y},p_{y}) \right]  \big\rbrace .
\end{split}
\end{equation}
The $ J_{L} $  is the propagator for the density matrix. To arrive at the Lindbladian terms in $ J_{L} $, use was made of equations (6) and (7). Note, the coordinates and momenta end points are the same as the end point conditions in equations (4) and (5). It is easy to see from equations (5) and (9) that both $  J  $ for a closed system and $ J_{L} $ for an open system with Markov conditions are Markov. 

Finally, consider an open system with memory term, for this I suggested an integro-differential equation, with the extra term as the Plus 
\begin{equation}\label{10}
i\hbar \dfrac{\partial \rho}{\partial t} =  H_{eff} \rho - \rho H^{\dagger}_{eff} + \textbf{L} \rho \textbf{L}^{\dagger} + \int_{0}^{t} d\tau \left[ M(t,\tau) \rho(\tau) - \rho(\tau)M^{\dagger}(t,\tau) + N(t,\tau)\rho(\tau)N^{\delta}(t,\tau) \right].
\end{equation}
In a previous paper \cite{Magpantay} , I showed that it gives a a propagator for the density matrix as given by 
\begin{equation}\label{11}
\begin{split}
J_{LP}(x,y; x',y'; t)&  = \int_{end points} (d\tilde{x}) (d\tilde{y}) (dp_{x}) (dp_{y})  \exp \frac{i}{\hbar} \int_{0}^{t} dt' \Big\lbrace \left[ p_{x}\cdot\dot{\tilde{x}} - H(\tilde{x},p_{x}) \right] \\
 &\quad - \left[ p_{y}\cdot \dot{\tilde{y}} - H(\tilde{y},p_{y}) \right] -i \left[ (\textbf{L}^{\dagger}\textbf{L})(\tilde{x},p_{x}) + (\textbf{L}^{\dagger}\textbf{L})(\tilde{y},p_{y}) + \textbf{L}(\tilde{x},p_{x})\textbf{L}^{\dagger}(\tilde{y},p_{y}) \right] \\
 & \quad + \int_{0}^{t'}  dt'' \Big[ M(\tilde{x},p_{x}; t',t'') - M^{\dagger}(\tilde{y},p_{y}; t',t'') + N(\tilde{x},p_{x}; t',t'') N^{\delta}(\tilde{y},p_{y}; t',t'') \Big\rbrace ,
 \end{split}
\end{equation}
 where the subscript LP in the above J stands for Lindblad Plus. The end points condition in equation (11) is the same as that in equation (9) and equation (5). 

Equations (1) to (11) summarize the relevant ideas and methods relevant to this problem. What they need is to derive the terms that appear in equations (9) and (11) from the underlying system, bath interaction. The Feynman-Vernon theory precisely provides the method for this. Unfortunately, the example they used and expounded on by Caldeira-Leggett \cite{Caldeira} only gives non-Markov terms. I will make use of their particular example and show that a Markov term, a Lindbladian can be derived also. I do this by modifying the particle-bath interaction they used. 

3. I will give first a summary of the (FV)/(CL) example. The system has a single particle with coordinate x and the bath is described by coordinates $ X_{i} $ with  $ i = 1,.., N $, N taken to be a large number. The bath is composed of independent harmonic oscillators each with a particular frequency $ \omega_{i} $ but all having the same mass m and the interaction of the particle, of mass M (use of capital M does not imply more massive than bath oscillators) with the bath, is given by a simple product term. The Lagrangians are
\begin{subequations}\label{12} 
\begin{gather}
L_{b} = \frac{m}{2} \sum_{i = 1}^{N} \left[ \dot{X}_{i}^{2} - \omega_{i}^{2} X_{i}^{2} \right] , \label{first}\\
L_{int} = -x\sum_{i = 1}^{N} c_{i}X_{i}, \label{second}\\
L_{s} = \frac{M}{2} \dot{x}^{2} - V(x),
\end{gather}
\end{subequations}
where $ c_{i} $ are constants, The interaction term says that all bath oscillators interact with the particle regardless of where the particle is. This is clearly unrealistic. 

THe Feynman-Vernon theory says the density matrix for the particle and bath is 
\begin{equation}\label{13}
\rho(x,X; y,Y; t) = \int dx'dX'dy'dY' K(x,X;x',X'; t) \rho(x',X'; y',Y'; 0) K^{*}(y',Y';y,Y; t),
\end{equation}
where the bath index i suppressed in the above equation for simplicity. Since the system, bath interaction is turned on only at $ t = 0 $, the initial density matrix factorizes  as
\begin{equation}\label{14}
\rho (x',X'; y',Y'; 0) = \rho_{s}(x',y'; 0) \rho_{b}(X',Y'; 0)
\end{equation}
where the subscripts s and b that goes with $ \rho $ refer to system and bath respectively. Focusing on the particle's density matrix, it is given by tracing over the bath's end points X and Y with X=Y, i.e.,
\begin{equation}\label{15}
\rho_{s}(x,y; t) = \int dX \rho(x,X;y,X; t).
\end{equation} 
Using equation (13) and (14) gives
\begin{subequations}\label{16}
\begin{gather}
\rho_{s}(x,y; t) = \int dx' dy' J(x,y;x',y'; t)_{FV} \rho_{s}(x',y'; 0), \label{first}\\
J_{FV}(x,y;x',y'; t) = \int dX dX' dY' K(x,X;x',X'; t) \rho_{b}(X',Y'; 0) K^{*}(y',Y';y,X; t),
\end{gather}
\end{subequations}
where the subscript FV in above J stands for Feynman-Vernon and integrations in the above J are over the bath end point conditions (note, the end point condition at t is only X because of the tracing made). Before expressing this $ J_{FV} $ in terms of a path-integral, it is important to notice that it hints it is not Markov because of the presence of the bath's $ \rho_{b}(X_{1},Y_{1}; 0) $ between the two K's. This should be expected because the bath oscillators incessant interaction with the system's degree of freedom (given by x) will imprint long term memory effects on the system. It is encouraging that the $ J_{FV} $ in equation (16b) has the same character, non-Markov, as the $ J_{LP} $ in the Lindblad Plus as given by equation (11). But in equation (11), it is explicitly constructed there is a bath effect that is Markov. Maybe it can also be present here.

The quantum propagator K for the system plus bath is given by
\begin{equation}\label{17}
\begin{split}
K(x,X;x',X'; t)& = \left\langle x X \Big\vert \exp {-\frac{i}{\hbar} H_{s+b} t} \Big\vert x'X' \right\rangle,\\
                         & = \int_{end points} (d\tilde{x}) (d\tilde{X}) \exp {\frac{i}{\hbar} \int_{0}^{t} dt' \left[ L_{s}(\tilde{x}) + L_{b}(\tilde{X}) + L_{int}(\tilde{x},\tilde{X}) \right] }.
\end{split}
\end{equation} 
The end points condition in above are $ \tilde{x}(t' = 0) = x', \tilde{x}(t' = t) = x, \tilde{X}(t' = 0) = X', \tilde{X}(t' = t) = X $, and there are similar conditions for $ \tilde{Y} $. Substituting equation (17) and a similar equation for $ K^{*} $ in equation (16b), the propagator for the distribution function of the system becomes
\begin{equation}\label{18}
J_{FV}(x,y;x',y'; t) = \int_{end points} (d\tilde{x}) (d\tilde{y}) \textbf{F}(\tilde{x},\tilde{y}) \exp {\frac{i}{\hbar}\int dt' \left[ L_{s}(\tilde{x}(t')) - L_{s}(\tilde{y}(t')) \right] }, 
\end{equation}
where the end points condition in above are $ \tilde{x}(t' = 0) = x' $, $ \tilde{x}(t' = t) = x $ and similar end point conditions for $ \tilde{y} $. The functional $ \textbf{F} $ is known as the influence functional and is given by the path-integral and ordinary integrals
\begin{equation}\label{19}
\begin{split}
\textbf{F}(\tilde{x},\tilde{y})& = 	\int dX dX' dY' \rho_{b}(X',Y'; 0) \int_{end points} (d\tilde{X}) (d\tilde{Y}) \exp \frac{i}{\hbar} \int dt' \Big\lbrace L_{b}(\tilde{X}(t')) + L_{int}(\tilde{x}(t'),\tilde{X}(t'))\\
&\quad - L_{b}(\tilde{y}(t')) -L_{int}(\tilde{y}(t'),\tilde{Y}(t')) \Big\rbrace 
\end{split}
\end{equation}
The end points in the above path-integral are defined by $ \tilde{X}(t' = 0) = X', \tilde{X}(t' = t) = X = \tilde{Y}(t' = t) , \tilde{Y}(t' = 0) = Y' $, and the ordinary integrals are over the initial and final end points. To carry out the evaluation of equation (19), I use a background decomposition
\begin{equation}\label{20}
\tilde{X}_{i} = \bar{X}_{i} + \delta X_{i},
\end{equation} 
where $ \bar{X}_{i} $ (I now use the index i to make clear there are many harmonic oscillators in the bath) is the harmonic oscillator state satisfying
\begin{equation}\label{21}
\dfrac{d^{2}\bar{X}_{i}}{dt'^{2}} + \omega_{i}^{2} \bar{X}_{i} = 0,
\end{equation}
and $ \delta X_{i} $ a fluctuation about this classical solution. The solution to equation (21) is 
\begin{equation}\label{22}
\bar{X}_{i}(t') = A_{i} \sin \omega_{i} t' + B_{i} \cos \omega_{i} t', 
\end{equation}
where the constants $ A_{i} $ and $ B_{i} $ are determined from the end point conditions 
$ \bar{X}_{i}(0) = X'_{i} $, $ \bar{X}_{i}(t) = X_{i} $ and gives
\begin{subequations}\label{23}
\begin{gather}
B_{i} = X'_{i}, \label{first}\\
A_{i} = X_{i} \dfrac{1}{\sin \omega_{i} t}  - X'_{i} \cot \omega_{i} t .
\end{gather}
\end{subequations}
The interaction term becomes
\begin{equation}\label{24}
S_{int} = \int_{0}^{t} dt' \tilde{x}(t') [ \bar{X}_{i}(t') + \delta X_{i}(t') ],
\end{equation}
and using equations (22) and (23)  will show how the end points of the bath oscillator appear with the particle's coordinate.

Since the end point conditions for $ \tilde{X}_{i} $ is accounted for by $ \bar{X}_{i} $, it means that the end point conditions for the fluctuations $ \delta X_{i}(t' = 0, or t) = 0 $.  
Substituting equation (20) in equations (12; a, b) to compute the influence functional $ \textbf{F}(\tilde{x},\tilde{y}) $ will involve the path integral of the fluctuations $ delta X_{i}(t') $ with a quadratic harmonic oscillator term and a linear term $ \tilde{x(t')} \cdot \delta X_{i}(t') $. The Greens function $ G(t',t'') $ of  $ \dfrac{d^{2}}{dt'^{2}} + \omega_{i}^{2} $ must be solved subject to the boundary conditions of the fluctuations and it is given by 
\begin{equation}\label{25}
G(t',t'') = \dfrac{1}{m \omega_{i} \sin \omega_{i} t } [\sin \omega_{i} (t - t') \sin \omega_{i} t'' ],
\end{equation}
valid for $ t'' \leq t' $ and t' in the range $ (0,t) $. 

The bath end points integrations (ordinary integrals), follow after (1) integration by parts of the kinetic and mass terms of $ \bar{X}_{i} $ and (2) making use of equations (22) and (23) in the first term of equation (24).

There are two more ingredients, (1) doing the same as above for the bath degrees of freedom $ \tilde{Y}_{i} $, with end point conditions $ \bar{Y}_{i}(t' = 0) = Y'_{i} $, $ \bar{Y}_{i}(t' = t) = X_{i} $, and (2) assuming that the bath degrees of freedom at $ t' = 0 $ is at thermal equilibrium with distribution
\begin{equation}\label{26}
\rho_{b}(X',Y'; 0) = \prod_{i} \dfrac{m\omega_{i}}{2\pi \hbar \sinh \frac{\hbar\omega_{i}}{kT}} \exp {\left\lbrace -\dfrac{m\omega_{i}}{2\hbar \sinh \frac{\hbar\omega_{i}}{kT}} \left[ (X_{i}^{\prime 2} + Y_{i}^{\prime 2}) \cosh \frac{\hbar \omega_{i}}{kT} - 2X'_{i}Y'_{i} \right] \right\rbrace }
\end{equation}     
The ordinary integrals on the bath end point conditions can be done to give
\begin{equation}\label{27}
\begin{split}
J_{FV}(x,y;x',y'; t)& = \int_{end points} (d\tilde{x})(d\tilde{y}) \exp  \frac{i}{\hbar} \big\lbrace  \int_{0}^{t} dt' \left[ L_{s}(\tilde{x},\dot{\tilde{x}}) - L_{s}(\tilde{y},\dot{\tilde{y}}) \right] \\ 
& \quad + \int_{0}^{t} dt' \int_{0}^{t'} dt'' [ \tilde{x}(t') - \tilde{y}(t') ] \left( i \sum_{i} \dfrac{c_{i}^{2}}{2m\omega_{i}} \left( \coth \frac{\hbar \omega_{i}}{2k T} \right) \cos \omega_{i}(t' - t'') \right) [ \tilde{x}(t'') - \tilde{y}(t'') ] \\
& \quad + \int_{0}^{t} dt' \int_{0}^{t'} dt'' [ \tilde{x}(t') - \tilde{y}(t') ] \left( \sum_{i} \frac{ c_{i}^{2}}{2m\omega_{i}} \sin \omega_{i}(t' - t'') \right) [ \tilde{x}(t'') + \tilde{y}(t'') ] \big\rbrace ,
\end{split}
\end{equation}
which is the same result given by Caldeira-Leggett. The imaginary number i that goes with the first sum becomes a decay term, gives the decoherence effect of the thermal initial state. Compairing with equation (11), the double time-integrals show that the bath effect on the particle's dynamics are purely non-Markov. They are all memory terms, which can be explained by the fact that the classical  harmonic oscillator solutions $ \bar{X}_{i} $ and the fluctuations $ \delta X_{i} $, all of them influence the particle coordinate $ \tilde{x} $ all the time, anywhere. Assuming no momentum dependence on the Markov and non-Markov terms of equation (11), comparison with equation (27) gives a zero Lindbladian and 
\begin{subequations}\label{28}
\begin{gather}
M(\tilde{x}; t',t'') = \tilde{x}(t') \sum_{i} \frac{c_{i}^{2}}{2m\omega_{i}} \left[ i \coth \frac{\hbar \omega_{i}}{2k T}\cos \omega_{i}(t'-t'') + \sin \omega_{i}(t'-t'') \right] \tilde{x}(t''),\label{second}\\
N(\tilde{x}; t',t'')N^{\delta}(\tilde{y}; t',t'') = \begin{pmatrix}
	                                                                                  A \tilde{x}(t') & B \tilde{x}(t'')
	                                                                                  \end{pmatrix}
	                                                                                  \cdot
	                                                                                  \begin{pmatrix}
	                                                                                  A \tilde{y}(t'')\\
	                                                                                  B \tilde{y}(t')
	                                                                                  \end{pmatrix},
\end{gather}
\end{subequations}	
with the values of A and B given by 
\begin{subequations}\label{29}
\begin{gather}
A^{2} = \sum_{i} \frac{c_{i}^{2}}{2m \omega_{i}} \left[ -i \coth \frac{\hbar \omega_{i}}{2k T} \cos \omega_{i}(t'-t'') + \sin \omega_{i}(t'-t'') \right] ,\label{first}\\
B^{2} =  \sum_{i} \frac{c_{i}^{2}}{2m \omega_{i}}\left[ -i \coth \frac{\hbar \omega_{i}}{2k T} \cos \omega_{i}(t'-t'') - \sin \omega_{i}(t'-t'') \right] 
\end{gather}
\end{subequations}
This is the reason that I stated in \cite{Magpantay} that the (FV) and (CL) example yielded ony the Lindblad Plus.

4. In this part I show how to get a Lindbladian from the (FV) and (CL) example. As I already pointed out in part 3, the interaction term equation (12 b) is unrealistic, the particle interacts with all oscillators all the time regardless of where the particle is.    
  
I assume that bath surrounds the particle with coordinate x, with the number of oscillators N being large and the particle needs equal time, call it $ \tau $, to bounce off from one bath oscillator to another, the collision time is infinitesimal and call it $ \epsilon $. It is clear that $ \epsilon $ is much less than $\tau $, the spacing between oscillators is much bigger than the size of the oscillators. I will label the particle's collision time sequentially as $ ( \tau_{1}, \tau_{2}, ...., \tau_{i},.... \tau_{N} = t ) $, where $ \tau_{i} = i\tau $ seconds is the time the ith oscillator collides with the particle, the first collision happens after $ \tau $ seconds from the start. I will also assume that the oscillators all have the same frequency $ \omega $, this will simplify the calculations and will be clear later, when I discuss the general case of many different frequencies. From equation (24), the interaction term becomes
\begin{equation}\label{30}
S_{int} = \sum_{i} \epsilon \tilde{x}(\tau_{i}) \left[ (\dfrac{-1}{\sin \omega t} X_{i} + \cot \omega t X'_{i}) \sin \omega \tau_{i} - X'_{i} \cos \omega \tau_{i} \right] + \int_{0}^{t} dt' \tilde{x}(t') \delta X_{i}(t'),
\end{equation}
the index i, which goes with time $ \tau_{i} $ is now the index for the end point conditions of the bath coordinates. This equation also says all the end point conditions are still determined at $ t' = 0 $ and at $ t' = t $, which is also used in part 4. But the ith oscillator only interacts with the particle during the collision at $ \tau_{i} $ for an infinitesimal time $ \epsilon $, and then this oscillator continues its oscillation without interacting any more with the particle, which by time $ \tau_{i+1} $ is now interacting with the $ i+ 1 $ oscillator. However, note the integral in equation (30), the weak fluctuation is assumed to interact throughout with the particle. 

Now in calculating the bath + particle propagator $ K^{*}(y',Y';y,X; t) $, the time interval (0,t) will also be divided by N $ \tau $ collision intervals. Since all oscillators have the same mass m and frequencies $ \omega $, we can lump the ith label with the collision time $ \tau_{i} $. Now follow the same calculation as before, the quadratic term in $ \delta X_{i}(t') $ and linear term with equation (30) will yield the same as equation (25). This will be the only double time integral in $ J_{FV} $,  memory term. The end point conditions $ X_{i}, X'_{i}, Y'_{i} $ that appear in $ \frac{m}{2}[ \bar{X}_{i}(t) \dot{\bar{X}_{i}}(t) - \bar{X}_{i}(0) \dot{\bar{X}_{i}}(0) ] $ and those that appear in an equation similar to the first term of (30) but now in terms of the end point conditions for $ Y $ will now be time ordered also by $ \tau_{i} $. When the end point ordinary integrals are done, it will involve end point conditions for the same time $ \tau_{i} $, thus the result involve only a single sum over time $ \tau_{i} $, which can be written as a single time integral. This is a Lindbladian. 

Now, the details. The quadratic kinetic term in $ \bar{X}_{i} $, which upon integration by parts gives $ \frac{m}{2}[ \bar{X}_{i}(t) \dot{\bar{X}_{i}}(t) - \bar{X}_{i}(0) \dot{\bar{X}_{i}}(0) ] $ is equal to
\begin{equation}\label{31}
QuadK =  \exp \frac{i}{\hbar} \sum_{i = 0}^{N} \Big\lbrace \frac{m\omega}{2} (\cot \omega t) (X_{i}^{2} + X'_{i}{2}) - \dfrac{m\omega}{\sin \omega t} X_{i}X'_{i},
\end{equation}  
and there will be a similar equation  for the $ Y $ terms, except has a $ \frac{-i}{\hbar} $. 
The linear in $ \bar{X}_{i} $ terms are given by
\begin{equation}\label{32}
Lin = \exp \frac{i}{\hbar} \sum_{i = 0}^{N} \epsilon c_{i }\tilde{x}(\tau_{i}) \left[ (\dfrac{-1}{\sin \omega t} X_{i} + \cot \omega t X'_{i}) \sin \omega \tau_{i} - X'_{i} \cos \omega \tau_{i} \right], 
\end{equation}
and there is a similar equation for the $ Y, y $ term. The fluctuation path-integral makes use of equation (25) and gives
\begin{equation}\label{33}
Fluct = \exp \frac{i}{\hbar} \sum_{i = 0}^{N} c_{i}^{2}  \frac{1}{m \omega} \dfrac{1}{\sin \omega t} \int_{0}^{t} dt' \int_{0}^{t'} dt'' \tilde{x}(t') \sin \omega (t - t') \sin \omega t'' \tilde{x}(t''), 
\end{equation}
and a similar expression involving $ \tilde{y} $. The index i that appear in (31) and (32) are connected with the time index $ \tau_{i} $, thus each end points integrals $ X_{i}, X'_{i}, Y'_{i} $ are done for each time and the final result is a single sum over $ \tau_{i} $. 
The end point integrals are simple. The terms suggest the $ X_{i} $ integral is done first, which will give a relationship between $ X'_{i} and Y'_{i} $. To get the final answer, do the final integral in terms of $ X'_{i} $ then go back to the intermediate step after the $ X_{i} $ integral, do the final integral with $ Y'_{i} $ and since the results should be equal, add the two results and divide by two to get the final answer for the generating functional J for the modified (FV)/(CL) example 
\begin{equation}\label{34}  
\begin{split}
J_{ME}(x,y;x',y'; t)& = \int_{end points} (d\tilde{x})(d\tilde{y}) \exp  \frac{i}{\hbar} \int_{0}^{t} \big\lbrace  dt' \left[ L_{s}(\tilde{x},\dot{\tilde{x}}) - L_{s}(\tilde{y},\dot{\tilde{y}}) \right] \\ & \quad + \begin{pmatrix} \\ \tilde{x}(t') & \tilde{x}(t') \end{pmatrix} \cdot \left[ \sum_{i} \dfrac{c_{i}^{2} \epsilon^{2}}{4m\omega} \left( -2 \sin 2\omega t' + \cot \omega t \sin^{2} \omega t' \right) \delta(t' - \tau_{i}) \right] \begin{pmatrix} 
                                                                                   \tilde{x}(t')\\
                                                                                   \tilde{x}(t')
                                                                                   \end{pmatrix} \\
& \quad +  \begin{pmatrix} \\ \tilde{y}(t') & \tilde{y}(t') \end{pmatrix} \cdot \left[ \sum_{i} \dfrac{c_{i}^{2} \epsilon^{2}}{4m\omega} \left( -2 \sin 2\omega t' + \cot \omega t \sin^{2} \omega t' \right) \delta(t' - \tau_{i}) \right] \begin{pmatrix} 
                                                                                \tilde{y}(t')\\
                                                                                \tilde{y}(t')
                                                                                \end{pmatrix} \\
& \quad +  \begin{pmatrix} \\ \tilde{x}(t') & \tilde{y}(t') \end{pmatrix} \cdot \left[ \sum_{i} \dfrac{c_{i}^{2} \epsilon^{2}}{4m\omega} \left( -2 \sin 2\omega t' + \cot \omega t \sin^{2} \omega t' \right) \delta(t' - \tau_{i}) \right] \begin{pmatrix} 
                                                                                \tilde{y}(t')\\
                                                                                \tilde{x}(t')
                                                                                \end{pmatrix} \big\rbrace \\                                                          
& \quad \exp  \frac{-1}{\hbar} \int_{0}^{t} dt' \big\lbrace \begin{pmatrix} \\ \tilde{x}(t') & \tilde{x}(t') \end{pmatrix} \cdot \left[ \sum_{i} \dfrac{c_{i}^{2} \epsilon^{2}}{4m\omega} \cos 2 \omega t' \coth\frac{\hbar \omega}{2k T} \delta(t' - \tau_{i}) \right] \begin{pmatrix}               
\tilde{x}(t')\\
\tilde{x}(t')
\end{pmatrix} \\
& \quad +  \begin{pmatrix} \\ \tilde{y}(t') & \tilde{y}(t') \end{pmatrix} \cdot \left[ \sum_{i} \dfrac{c_{i}^{2} \epsilon^{2}}{4m\omega} \cos 2 \omega t'  \coth\frac{\hbar \omega}{2k T} \delta(t' - \tau_{i}) \right] \begin{pmatrix} 
                                            \tilde{y}(t')\\
                                            \tilde{y}(t')
                                            \end{pmatrix} \\
& \quad +  \begin{pmatrix} \\ \tilde{x}(t') & \tilde{y}(t') \end{pmatrix} \cdot \left[ \sum_{i} \dfrac{c_{i}^{2} \epsilon^{2}}{4m\omega} \cos 2 \omega t' \coth\frac{\hbar \omega}{2k T} \delta(t' - \tau_{i}) \right] \begin{pmatrix} 
                                        \tilde{y}(t')\\
                                        -\tilde{x}(t')
                                        \end{pmatrix} \big\rbrace \\                                                                                                                             
& \quad + exp \frac{i}{\hbar} \sum_{i = 0}^{N} \dfrac{c_{i}^{2}}{m\omega \sin \omega t} \int_{0}^{t} dt' \int_{0}^{t'} dt'' \begin{pmatrix} \\ \tilde{x}(t') & \tilde{y}(t') \end{pmatrix} \cdot \sin \omega (t - t') \sin \omega t'' \begin{pmatrix} 
                                                               \tilde{x}(t')\\
                                                               -\tilde{y}(t'')
                                                               \end{pmatrix}.  
\end{split}
\end{equation}
Equation (34) clearly shows the single time integral characteristic of a LIndbladian, which is read off by comparing with equation (11) under the no momentum dependence. This Lindbladian has a time-decoherence effect due to the initial thermal distribution of the bath, same as the (FV)/(CL) example but there coming from the purely non-Markov effect. The non-Markov terms of equation (35) is quite simple because it comes about from the fluctuation terms of the bath harmonic oscillator  interacting incessantly with the particle coordinate $ \tilde{x}(t') $. In the (FV)/(CL) example, even the classical harmonic oscillator  
interacts all the time and all at the same time with the particle. 

5. Now suppose the oscillators have different frequencies $ (\omega_{1}, ...., \omega_{k}, ...) $, and that the particle only interacts with an oscillator if it is around or near the vicinity of that oscillator of frequency $ \omega_{k} $ in time $ \epsilon $ then moves on to the next oscillator in time $ \tau $. The calculation will still show a Lindbladian because the calculation done in part 4 still can be done, to a certain extent. Using equations (20), (22) (23) in the first term equation (24), the integral can made into a sum over time $ (\tau_{1}, \tau_{2}, ..., \tau_{i}, ...) $ where the index i is the same as the index in the frequencies $ \omega_{i} $ when following the evolution from bath oscillators $ \tilde{X_{i}}(t'=0) = X'_{i} $ to  $ \tilde{X_{i}}(t'=t) = X_{i} $. But when we follow the the evolution from bath oscillators $ \tilde{Y_{k}}(t'=0) = Y'_{k} $ to  $ \tilde{Y_{k}}(t'=t) = Y_{k} $, the time assignments $ (\tau_{1}, \tau_{2}, .... \tau_{i}, ...) $ will not have the same indices that goes with the frequencies and end points. Still the sequential time assignments can be made but for a given time $ \tau_{i} $ the end point conditions that appear and the frequencies will have two different indices. For that particular time $ \tau_{i} $, we may have $ X_{i}, X'_{i}, Y'_{i} $ and $ X_{k}, X'_{k}, Y'_{k} $. But still the end point integrals can be done two at a time but more difficult and the results will not be as nice and simple as in part 4. In other words, the very local interaction of the particle with the oscillators will still yield a Lindbladian, The non-Markov term will still be the same as in part 4.

6. There are many Lindblad systems discussed in the literature, for examples, see \cite{Chruscinski}. These examples can be called phenomenological. It would be nice if a derivation from a system interacting with an environment can be shown for them. Maybe, the interaction can give another aspect, i.e., they may have a non-Markov contribution. 

In conclusion, it is nice to find that the two methods, the path-integral or sum over histories by Feynman-Vernon and the (LGKS), in dealing with quantum statistical mechanics can be related and are consistent.

\begin{acknowledgments}
I have always wanted to find out how to derive (LGKS) from (FV). In learning this, Gravity, a loyal company, kept me focused and distracted, both needed for me to keep on working. Thank you Gravity. 
\end{acknowledgments}

\bibliography{apssamp}

\end{document}